\documentclass[aps,pra,twocolumn,superscriptaddress,notitlepage,nofootinbib,longbibliography, colorlinks=true]{revtex4-2}
\usepackage{amsmath,amssymb,amsfonts,graphicx,float,times,psfrag}
\usepackage[pdftex]{color}
\usepackage{amsmath,bm}
\usepackage[colorlinks, linkcolor=blue, citecolor=blue,  breaklinks=true]{hyperref}
\usepackage{mathtools,amsfonts,mathptmx}
\usepackage[utf8]{inputenc}
\usepackage[T1]{fontenc}
\usepackage{xcolor}
\usepackage{braket}
\usepackage[titletoc,title]{appendix}
\usepackage[group-minimum-digits=4,
            per-mode=symbol,
            separate-uncertainty=true,
            range-phrase=--,
            detect-weight=true,
            detect-family=true]{siunitx}
\usepackage[capitalize,noabbrev]{cleveref}
\Crefname{equation}{Eq.}{Eqs.}
\crefname{equation}{Equation}{Equations}

\usepackage[shortlabels]{enumitem}
\begin{document}

\title{ Squeezing enhanced nonreciprocal quantum correlations via Barnett effect}

\author{E. \surname{Kongkui Berinyuy}}
\email{emale.kongkui@facsciences-uy1.cm}
\affiliation{Department of Physics, Faculty of Science, University of Yaounde I, P.O.Box 812, Yaounde, Cameroon}

\author{A.-H. Abdel-Aty}
\affiliation{Department of Physics, College of Sciences, University of Bisha, Bisha 61922, Saudi Arabia}

\author{P. Djorwé}
\email{djorwepp@gmail.com}
\affiliation{Department of Physics, Faculty of Science,
University of Ngaoundere, P.O. Box 454, Ngaoundere, Cameroon}

\author{N. Alessa}
\email{naalessa@pnu.edu.sa}
\affiliation{Department of Mathematical Sciences, College of Science, Princess Nourah bint Abdulrahman University, P.O. Box 84428, Riyadh 11671, Saudi Arabia}

\author{K.S. Nisar}
\email{n.sooppy@psau.edu.sa}
\affiliation{Department of Mathematics, College of Science and Humanities in Al-Kharj, Prince Sattam Bin Abdulaziz University, Al-Kharj 11942, Saudi Arabia}

\begin{abstract}
Cavity optomagnonic platforms offer a promising route for exploring quantum phenomena, particularly quantum correlations, which are vital resources for modern quantum technologies. Here, we propose a theoretical scheme for achieving nonreciprocal quantum correlations such as entanglement, and quantum discord via Barnett effect in a molecular-optomagnonical system, where a yttrium iron garnet sphere is placed in a microwave cavity that is hosting molecules. We show optimal parameter regimes for achieving nonreciprocal quantum correlations through Barnett effect. The generated entanglements are robust against thermal fluctuations, persisting even at high temperatures. Our scheme suggests a new tool for engineering noise-tolerant quantum correlations, and paves a way toward realizing novel nonreciprocal quantum devices by integrating magnons with molecular ensembles.
\end{abstract}

\maketitle

\section{Introduction} \label{sec:Intro}
In magnetic crystals, the magnetic moments are generally not isolated. Their mutual interactions lead to collective excitations, referred to as spin waves (magnons)~\cite{stancil2021}, which are commonly observed in ferrimagnetic materials such as yttrium iron garnet (YIG) sphere. Magnons have attracted significant attention due to their ability to couple strongly with microwave fields and various quantum systems. Interestingly, magnons can interact with lattice vibrations (phonons) via the magnetostrictive force~\cite{Potts2021}, giving rise to cavity magnomechanics (CMM). This interaction enables the exploration of diverse phenomena and applications, including entanglement~\cite{Hussain2022,Qiu2022,Sohail:23}, quantum steering~\cite{Ge2025}, ultrasensitive sensing~\cite{Colombano2020}, magnon-squeezing states~\cite{Li2019,Lu2023} and ultraslow light propagation~\cite{Lu2023}.

In recent years quantum correlations such as entanglement, and quantum discord has been investigated in cavity optomechanical (COM)~\cite{Huang2022,Tchounda2023,Massembele2025,Sohail2025,Bemani2019,Shang2024,Ge2025,Massembele2024,Emale2025,Djo2024,Chen2025,Emale2025}, molecular optomechanical (McOM)~\cite{Huang2024,Berinyuy2025a,Berinyuy2025Q,BERINYUY2026117820}, and cavity magnomechanical systems \cite{Li2018}. In particular, 
 nonreciprocal entanglement in cavity optomechanical (COM) \cite{Jiao2022,Jiao.2020}, molecular optomechanical (McOM)~\cite{BERINYUY2025}, and cavity magnomechanical (CMM) systems \cite{Chen2023} has attracted great interest. The ability to engineer nonreciprocal entanglement in these systems not only deepens our understanding of hybrid quantum systems but also opens up promising avenues for quantum information processing and plethora of quantum computational tasks~\cite{Xu2020}. While the Sagnac effect and chiral coupling~\cite{Lu2024} have been used to realize nonreciprocity in COM systems, mechanisms such as magnon Kerr nonlinearity~\cite{Chen2024} and  Barnett effect~\cite{Lu2025} induce similar behaviour in CMM systems. This Barnett effect, discovered in 1915~\cite{ Barnett1915}, is a phenomenon where the rotation of an object with magnetic moments leads to magnetization. Nevertheless, the possibility of achieving nonreciprocal bipartite, tripartite entanglement, and quantum discord via the Barnett effect in a hybrid magnon-molecular system with magnon squeezing remains unexplored and will be the subject of our research.

In this work, we propose a theoretical scheme to generate and control nonreciprocal bipartite and tripartite entanglement, and quantum discord via the Barnett effect within a hybrid magnon-molecular system with magnon squeezing. This configuration features a spinning YIG sphere and ensemble of molecules placed in  an optical cavity. The Barnett frequency shift of the magnon mode is induced by the rotation of the YIG sphere. This Barnett frequency shift can be turned from positive to negative by adjusting the direction of the rotation or bias magnetic field. This results in the generation of unidirectional entanglement, and quantum discord. The underlying physical mechanism is that, due to the angular momentum conservation, the Barnett effect reverses the polarity of the induced magnetization when the bias magnetic field or the rotation direction is reversed, resulting in time-reversal symmetry breaking in the magnon-molecular system. Furthermore, our findings strikingly unveil a thermal robustness of our proposed magnon-molecular system. It maintains significant quantum correlations at high temperatures. Finally, we demonstrate that perfect nonreciprocal bipartite and tripartite entanglement can be generated and controlled by exploring bidirectional contrast ratio. Our work paves a way for realizing novel nonreciprocal devices, and suggests a tool for engineering noise-tolerant quantum correlations.

The rest of our paper is structured as follows: \Cref{sec:model} introduces the theoretical model and outlines the derivation of the dynamical equations. \Cref{sec:results} delves into the numerical results and offers a thorough discussion of the underlying quantum correlations. Finally, \Cref{sec:concl} provides concluding remarks.

\section{Theoretical Model and dynamical equations} \label{sec:model}

\begin{figure*}[htp!]
	\centering
	\includegraphics[width=1\linewidth]{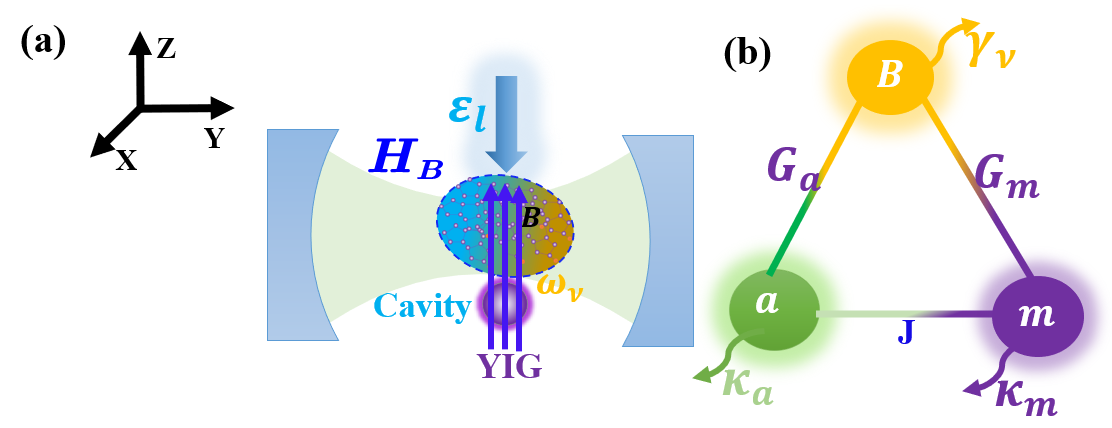}
	\caption{ (a) A magnon-molecular configuration featuring an ensemble of $N$ identical molecules  placed within a microwave cavity and a YIG sphere, with the magnon mode fully magnetized by the bias magnetic field $\mathbf{H}_0$ (not shown). The rotating YIG sphere, with angular frequency $\Delta_B$  generates an emergent magnetic field $\mathbf{H}_B$ that induces a frequency shift in the magnon. (b) A schematic of the interaction picture illustrating the cavity-magnomechanical coupling between the cavity mode ($a$) and magnon mode $m$ as well as coupling between the magnon mode $m$ and the collective molecular vibrational mode ($B$).}
	\label{fig:setup}
\end{figure*}

 Our benchmark system consists of a microwave cavity with frequency $\omega_a$ that hosts a YIG sphere supporting a magnon mode with frequency $\omega_m$ and a molecular mode with frequency $\omega_\nu$ as  sketched in \Cref{fig:setup}. Applying a bias magnetic field $\mathbf{H}_0$ along the z-axis induces a uniform magnon mode in the sphere at the resonance frequency $\omega_m=\gamma\mathbf{H}_0$, with $\gamma$ the gyromagnetic ratio.  The rotating YIG sphere, with angular frequency $\Delta_B$  generates an emergent magnetic field $\mathbf{H}_B=\Delta_B/\gamma$ that induces a frequency shift in the magnon.

 The Hamiltonian of the proposed system in the frame rotating at the driving frequency reads ($\hbar=1$)
\begin{equation}
\begin{aligned}
H=&\Delta_a a^\dagger a+(\Delta_m+\Delta_B)m^\dagger m+\sum_{j=1}^N \omega_\nu b_j^\dagger b_j+J(a^\dagger m+am^\dagger)\\&+\sum_{j=1}^N g_a a^\dagger a(b_j^\dagger + b_j)+\sum_{j=1}^N g_m m^\dagger m(b_j^\dagger + b_j)
+ i\mathcal{E}_l(m^\dagger + m)\\&+\frac{i}{2}\chi m^{\dagger 2}e^{i\theta}-H.c.,
\end{aligned}
\end{equation}
where  $a (a^\dagger)$, $m (m^\dagger)$, and $b_j (b_j^\dagger)$ are the annihilation (creation) operators for the cavity mode $a$, magnon $m$, and the $j$-th individual molecular vibrational mode, respectively. The quantities $\Delta_a=\omega_a-\omega_l$ and  $\Delta_m=\omega_m-\omega_l$  represent the frequency detuning of the cavity mode $a$ and for magnon mode $m$. The magnon-photon coupling rate is captured by $J$,  and $g_a$ ($g_m$) denotes the molecular-photon (molecular-magnon) coupling constant. The quantity $\frac{i}{2}\chi m^{\dagger 2}e^{i\theta}-Hc.
	$ captures the squeezing of the magnon mode with squeezing parameter $\chi$ and phase $\theta$. This can be realized by transferring squeezing from the cavity with squeezed vacuum field~\cite{Li2019,Lu2023}. The quantity $\mathcal{E}_l=\gamma\sqrt{5\mathcal{N}}B_0/4$ denotes the coupling strength of the driving magnetic field with amplitude $B_0$, and $\mathcal{N}=\rho V$ is the total number of spins, where $\rho=\num{4.22e27}m^{-3}$. Here, $\rho$ is the spin density and $V$ is the volume of the YIG sphere \cite{Shen2022,Li2018,Potts2021}.

The above Hamiltonian can be simplified by introducing the molecular collective operator, 
$
B=\sum_{j=1}^{N}b_j/N,$ where $[B,B^\dagger]=1$,

and our Hamiltonian can be rewritten as, 
\begin{equation}\label{eq:3}
\begin{aligned}
H=&\Delta_a a^\dagger a+(\Delta_m+\Delta_B)m^\dagger m+\omega_\nu B^\dagger B+G_aa^\dagger a(B^\dagger + B)\\&+G_mm^\dagger m(B^\dagger + B)+J(a^\dagger m+am^\dagger)+ i\mathcal{E}_l(m^\dagger + m)\\&+\frac{i}{2}\chi m^{\dagger 2}e^{i\theta}-H.c.,
\end{aligned}
\end{equation}
where $G_a=g_a\sqrt{N}$ and $G_m=g_m\sqrt{N}$ are the collective optomechanical coupling strengths.

\subsection{Quantum Langevin Equations and linearization}\label{sec:qle}
The evolution of the system in question is described by the set of quantum Langevin equations (QLEs),
 \begin{equation}\label{QLE}
\begin{aligned}
\dot{a}=&-(i\Delta_{a}+\kappa_a)a-iG_aa(B^\dagger+B)-iJm+\sqrt{2\kappa_a}a^{in},\\
\dot{m}=&-(i(\Delta_m+\Delta_B)+\kappa_m)m-iGm(B+B^\dagger)-iJa+\mathcal{E}_l\\&+\chi e^{i\theta}m^\dagger+\sqrt{2\kappa_m}m^{in},\\
\dot{B}=&-(i\omega_\nu+\gamma_\nu)B-iG_aa^\dagger a-iG_mm^\dagger m+\sqrt{2\gamma_\nu}B^{in}.
\end{aligned}
\end{equation}
The noise operators $a^{\text{in}}$, $m^{\text{in}}$, and $B^{\text{in}}$, where  $B^{\text{in}} = \frac{1}{\sqrt{N}}\sum_{j=1}^N b^{\text{in}}_j$  have zero mean values and are characterized by their following correlation functions,
\begin{equation}\label{eq:5}
\begin{aligned}
&\langle a^{\text{in}}(t)a^{\text{in}\dagger}(t^\prime)\rangle = \delta(t-t^\prime), 
\langle m^{\text{in}}(t)m^{\text{in}\dagger}(t^\prime)\rangle = \delta(t-t^\prime), \\
&\langle B^{\text{in}}(t)B^{\text{in}\dagger}(t^\prime)\rangle = (n_{\text{th}} + 1)\delta(t-t^\prime),\\& 
\langle B^{\text{in}\dagger}(t)B^{\text{in}}(t^\prime)\rangle = n_{\text{th}}\delta(t-t^\prime),
\end{aligned}
\end{equation}
where $n_{\text{th}} = \left[\exp\left(\hbar\omega_\nu/k_B T\right) - 1\right]^{-1}$ is the thermal phonon number for the mechanical mode at temperature $T$, and $k_B$ is the Boltzmann constant.
 Considering a strong drive, the dynamics of our system can be linearized by splitting each operator into the sum of its steady-state mean value and a small fluctuation around it, i.e $\mathcal{O}=\langle \mathcal{O}\rangle +\delta \mathcal{O} $, where $\langle \mathcal{O}\rangle\equiv(\alpha,m_s,\beta)$ stands for the steady-state mean value and $\delta \mathcal{O}$ captures the quantum fluctuation. Therefore, the linearized QLEs read,
\begin{equation}\label{fluc}
\begin{aligned}
\delta\dot{a}&=-(i\tilde{\Delta}_a+\kappa_a)\delta a-iJ\delta m+\sqrt{2\kappa_a}a^{in},\\
\delta\dot{m}&=-(i(\tilde{\Delta}_m+\Delta_B)+\kappa_m)\delta m-i\tilde{G}(\delta B+\delta B^\dagger)-iJ\delta a\\&+\chi e^{i\theta}\delta m^\dagger+\sqrt{2\kappa_m}m^{in},\\
\delta\dot{B}&=-(i\omega_\nu+\gamma_\nu)\delta B-i\tilde{G}(\delta m^\dagger+\delta m)+\sqrt{2\gamma_\nu}B^{in},
\end{aligned}
\end{equation}
where  $\kappa_a$, $\kappa_m$ and $\gamma_\nu$ are the decay rates of the cavity mode, magnon mode and collective vibrational mode, respectively. We have also defined the following effective detunings $\tilde{\Delta}_a=\Delta_{a}+2G_a\text{Re}[\beta]$ and $\tilde{\Delta}_m=\Delta_{m}+2G_m\text{Re}[\beta]$, together with their related effective couplings $\tilde{G}_a=g_a|\alpha|$ and $\tilde{G}_m=g_m|m_s|$. In this work, $\alpha$ and $m_s$ are assumed to be a real numbers, as we can properly adjust the phase reference of the cavity fields. Similarly, the steady-state mean valued equations yield,
\begin{equation}
\begin{aligned}
{\alpha}&=\frac{-iJm_s}{(i\tilde{\Delta}_a+\kappa_a)},~~
m_s=\frac{\mathcal{E}_l-iJ\alpha}{(i(\tilde{\Delta}_m+\Delta_B)+\kappa_m-\chi e^{i\theta})},\\
{\beta}&=\frac{-iG_a|\alpha|^2-iG_m|m_s|^2}{i\omega_\nu+\gamma_\nu}.
\end{aligned}
\end{equation}

To quantify quantum correlations, we define the  following quantum quadratures operators,
 $\delta X_o=\frac{( \delta o+\delta o^{\dagger})}{\sqrt{2}},~~~ \delta Y_o=\frac{( \delta o- \delta o^{\dagger})}{i\sqrt{2}}$ with $(o\equiv a, m,B)$, together with their corresponding noise operators, i.e.,  $X_o^{\text{in}}=\frac{(o^{\text{in}}+ o^{\text{\text{in}}\dagger})}{\sqrt{2}},~ Y_o^{in}=\frac{( o^{\text{\text{in}}}- o^{in\dagger})}{i\sqrt{2}}$. Therefore, our fluctuation dynamics in \Cref{fluc} can be written in a compact form as, 
\begin{equation}\label{eq:4}
\dot{\bm{u}}(t) = A\bm{u}(t) + \bm{n}(t),
\end{equation} 
with $\bm{u}^\top= \left( \delta X_a, \delta Y_a, \delta X_m, \delta Y_m, \delta X_B, \delta Y_B \right)$, and   
$\bm{n}^\top=( \sqrt{2\kappa_a}X_a^{\text{in}}, \sqrt{2\kappa_a}Y_a^{\text{in}}, \sqrt{2\kappa_c}X_m^{\text{in}}, \sqrt{2\kappa_c}Y_m^{\text{in}}, \sqrt{2\gamma_m}X_B^{\text{in}}, \sqrt{2\gamma_m}Y_B^{\text{in}})$. The drift matrix $\rm A$ reads,
\begin{widetext}
\begin{equation}
\rm A=
\begin{pmatrix}
-\kappa_a & \tilde{\Delta}_a & 0  & J & 0 & 0 \\
-\tilde{\Delta}_a & -\kappa_a & -J  & 0 & -2\tilde{G}_a & 0 \\
0 & J  & -\kappa_m+\chi\cos\theta & \tilde{\Delta}+\chi\sin\theta & 0 & 0 \\
- J  & 0 & -\tilde{\Delta}+\chi\sin\theta & -\kappa_m-\chi\cos\theta & -2 \tilde{G}_m & 0 \\
0 & 0 & 0 & 0 & -\gamma_\nu & \omega_\nu \\
-2\tilde{G}_a & 0 & -2 \tilde{G}_m & 0 & -\omega_\nu & -\gamma_\nu
\end{pmatrix},
\end{equation}
\end{widetext}
where $\tilde{\Delta}=(\tilde{\Delta}_m+\Delta_B)$. For our system to be stable, all real parts of the eigenvalues of the drift matrix $\rm A$ must be negative. This essential stability condition is achieved using the Routh-Hurwitz criterion~\cite{DeJesus1987}.
At this juncture, we introduce the covariance matrix $\rm V$, in order to explore the steady-state quantum correlations
 $\bm{V}_{lk} = \frac12\left\{ \langle \bm{u}_\ell(t)\bm{u}_k(t^{\prime}) \rangle + \langle \bm{u}_k(t^{\prime})\bm{u}_\ell(t) \rangle \right\}$. Equivalently, the elements of the covariance matrix $\rm V$ can be also derived by numerically solving the following Lyapunov equation \cite{Gardiner2000,Vidal2002},
\begin{equation}\label{eq:lyapunov}
A\bm{V} + \bm{V}A^\top = -\bm{D},
\end{equation}
where the diffusion matrix $\bm{D}$, which accounts for the influence of the noise operators, is given by,
\begin{equation}\label{eq:diffusion_matrix_D}
\bm{D} = \text{diag}\left[ \kappa_a, \kappa_a, \kappa_m, \kappa_m, \gamma_\nu(2n_{\text{th}} + 1), \gamma_\nu(2n_{\text{th}} + 1) \right].
\end{equation}

\subsection{Quantification of Bipartite and tripartite Entanglements}\label{sec:entanglement}

We employ the logarithmic negativity, $E_n$ to quantify bipartite entanglement in our system. This is a well-established entanglement monotone for Gaussian states~\cite{Plenio2005}. Its mathematical expression is given by,
\begin{equation}\label{eq:log_neg_En}
E_n = \max\left[0, -\ln(2\zeta)\right],
\end{equation}
where,
\begin{equation}\label{eq:log_neg_Gamma}
\zeta \equiv \frac{1}{\sqrt2} \sqrt{\Sigma(V) - \sqrt{\Sigma(V)^2 - 4\det(V)}},
\end{equation}
with $\Sigma(V) = \det(\varphi_1) + \det(\varphi_2) - 2\det(\varphi_3)$. For any given bipartite subsystem, the  $4\times4$ submatrix $V_{\text{sub}}$ extracted from the full $6\times6$ CM takes the canonical block form,
\begin{equation}\label{eq:V_sub_block_form}
V_{\text{sub}} =
\begin{pmatrix}
\varphi_1 & \varphi_3 \\
\varphi_3^\top & \varphi_2 \\
\end{pmatrix},
\end{equation}
where $\varphi_1$ and $\varphi_2$ are $2\times2$  matrices representing each subsystem, and $\varphi_3$ is a $2\times2$ matrix encoding the correlations  between the subsystems.

The minimum residual contangle $\mathcal{R}_{\tau}^{min}$ is used to quantify tripartite quantum entanglement. It is define as \cite{Adesso2006,Adesso2007}, 
\begin{equation}
\mathcal{R}_{\tau}^{min}\equiv \text{min}\left[\mathcal{R}_{\tau}^{a|mB},\mathcal{R}_{\tau}^{m|aB},\mathcal{R}_{\tau}^{B|am}\right].
\end{equation}
This expression guarantees the invariance of tripartite entanglement under all possible permutations of the modes $\mathcal{R}_{\tau}^{r|st}\equiv \mathcal{C}_{r|st}-\mathcal{C}_{r|s}-\mathcal{C}_{r|t},~(r,s,t\equiv a,m,B)$, and it satisfies the monogamy of quantum entanglement $\mathcal{R}_{\tau}^{r|st}\geq 0$.
Here, $\mathcal{C}_{u|v}$ is the contangle of subsystem $u$ and $v$ ($v$ contains one or two modes), which can be defined as squared logarithmic negativity and it is a proper entanglement monotone i.e., $\mathcal{C}_{u|v}=E^2_{N_{u|v}}$, with 
$E_{N}\equiv \text{max}\left[0,-\ln2\eta\right]$
, where 
$\eta=\text{min}~ \text{eig}[i\Omega_{3}\eta_{6}^\prime]$, with $\Omega_{3}$ and $\eta_{6}^\prime$ defined respectively as,
\begin{equation}
\Omega_{3}=\bigoplus_{k=1}^3i\sigma_{y},~~ \sigma_{y}=\begin{pmatrix}
0&-i\\i&0
\end{pmatrix}
\end{equation} 
and, 
$\eta^\prime_{6}=P_{r|st}V_{6}P_{r|st}~~ \text{for}~~ r,s,t\equiv a,m,B$, where 
$P_{a|mB}=\text{diag(1,-1,1,1,1,1)}$, $P_{m|aB}=\text{diag(1,1,1,-1,1,1)}$, and $ P_{B|am}=\text{diag(1,1,1,1,1,-1)}$ are partial transposition matrices and $V_6$ is $6\times 6$ CM of the three modes.

\subsection{Gaussian Quantum Discord (GQD)}\label{sec:discord}

In order to realize a more comprehensive understanding of quantum correlations, we explore the Gaussian quantum discord (GQD). This quantity captures all forms of nonclassical correlations including those that exist even in separable states making it a valuable tool for quantum information tasks where entanglement alone is not present~\cite{Ollivier2001,Giorda2010}. Consider a bipartite Gaussian subsystem $l$, characterized by its $4\times4$ covariance matrix $V_{\text{sub}}$, the GQD is given by~\cite{Giorda2010},
\begin{equation}\label{eq:GQD_DGl}
D_G^\ell = g(\sqrt{I_1^\ell}) - g(v^\ell_-) - g(v^\ell_+) + g(\sqrt{\mathcal{W}^\ell}),
\end{equation}
where the function $g(x)$ is defined as,
\begin{equation}\label{eq:GQD_f_x}
g(x) = \left(x + \frac12\right)\ln\left(x + \frac12\right) - \left(x - \frac12\right)\ln\left(x - \frac12\right).
\end{equation}
The symplectic eigenvalues $v^\ell_-$ and $v^\ell_+$ of the subsystem's covariance matrix are obtained from,
\begin{equation}\label{eq:GQD_v_pm}
v^\ell_{\pm} \equiv \frac{1}{\sqrt2} \left\{ \Sigma(V^\ell_{\text{sub}}) \pm \left[ \Sigma(V^\ell_{\text{sub}})^2 - 4I^\ell_4 \right]^{1/2} \right\}^{1/2},
\end{equation}
with $\Sigma(V_{\text{sub}}) = I^\ell_1 + I^\ell_2 + 2I^\ell_3$. Here, $I^\ell_1 = \det(\varpi_1)$, $I^\ell_2 = \det(\varphi_2)$, $I^\ell_3 = \det(\varphi_3)$, and $I^\ell_4 = \det(V_{\text{sub}})$. The term $\mathcal{W}^\ell$ is an important quantity that characterizes the quantumness of correlations and is given by~\cite{Chakraborty2017},
\begin{widetext}
	\begin{equation}\label{eq:GQD_W_ell}
	\mathcal{W}^\ell =
	\begin{cases}
	\left( \frac{2|I^\ell_3| + \sqrt{4I_3^{l2} + (4I^\ell_1 - 1)(4I^\ell_4 - I^\ell_2)}}{4I^\ell_1 - 1} \right)^2 & \text{if} \quad \frac{4(I^\ell_1I^\ell_2 - I^\ell_4)^2}{(I^\ell_2 + 4I^\ell_4)(1 + 4I^\ell_1)I^{l2}_3} \leq 1, \\
	\frac{I^\ell_1I^\ell_2 + I^\ell_4 - I^{l2}_3 - \sqrt{(I^\ell_1I^\ell_2 + I^\ell_4 - I^{l2}_3)^2 - 4I^\ell_1I^\ell_2I^\ell_4}}{2I^\ell_1} & \text{otherwise}.
	\end{cases}
	\end{equation}
\end{widetext}
An essential feature of the Gaussian states is that they exhibit entanglement if and only if their smallest symplectic eigenvalue $v_-$ is less than $1/2$. This shows that while $0 \leq D^\ell_G \leq 1$ can correspond to separable or entangled states, a value of $D^\ell_G > 1$ definitively indicates an entangled state~\cite{Giorda2010}.

\section{Results}\label{sec:results}

In our numerical simulation, we have used the following experimentally feasible parameter  \cite{KoczorBenda2022,Chikkaraddy2022,Zou2024,Roelli2024}, i.e., $\omega_\nu/2\pi = \SI{30}{\tera\hertz}$, $\mathcal{E}_l/\omega_\nu = 3.8$, $\gamma_\nu/\omega_\nu= \num{0.005}$, $g_\nu/\omega_\nu = \num{3.3e-6}$, $\Delta_m/\omega_\nu=1$, $\Delta_a/\omega_\nu=-1$, $|\Delta_B|/\omega_\nu=0.3$, $g_a=g_\nu$, $g_m=g_\nu$, $\kappa_a/\omega_\nu=0.0166$, $\kappa_a= \kappa_m$, $J/\omega_\nu = \num{0.2}$, and $N = \num{e7}$, and $T = \SI{210}{\kelvin}$.
\begin{figure}[tbh]
	\centering
	\includegraphics[width=4.2cm]{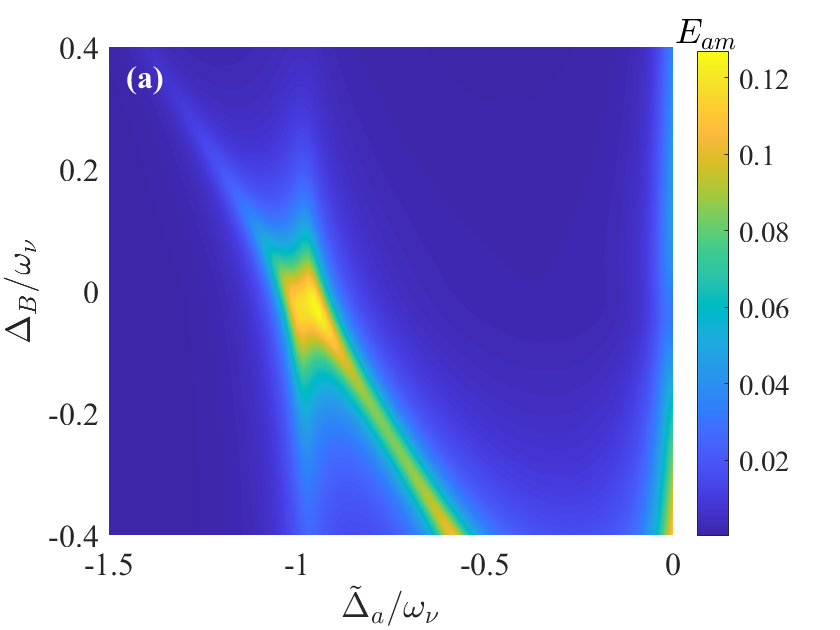}
	\includegraphics[width=4.2cm]{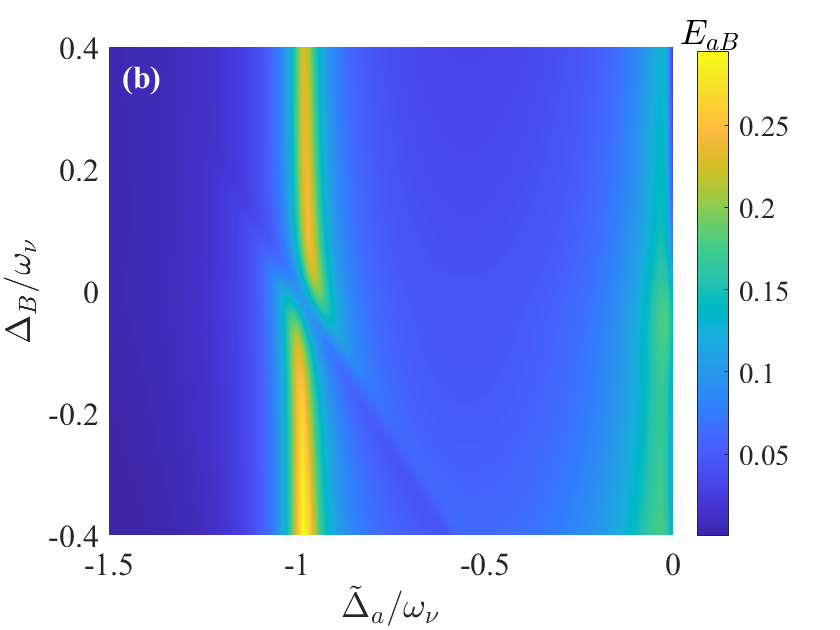}
	\includegraphics[width=4.2cm]{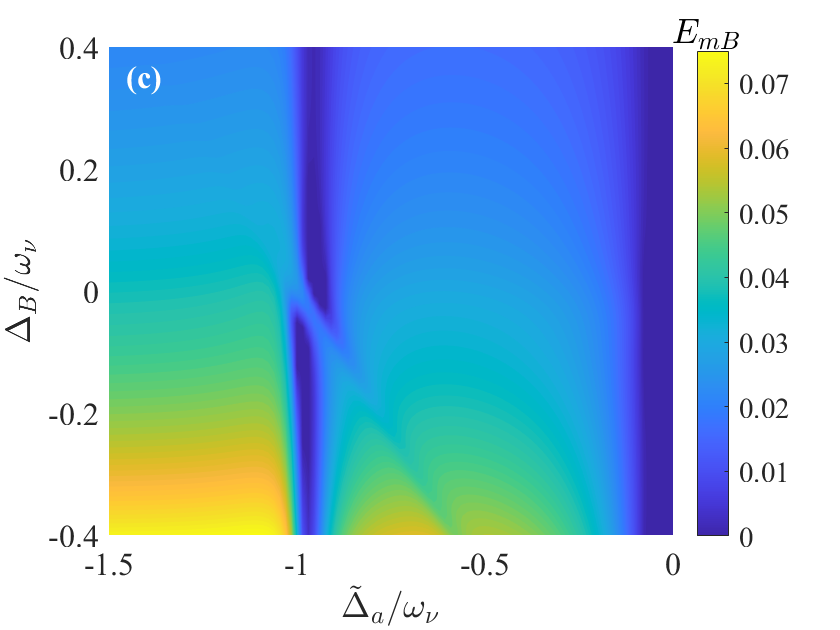}
	\includegraphics[width=4.2cm]{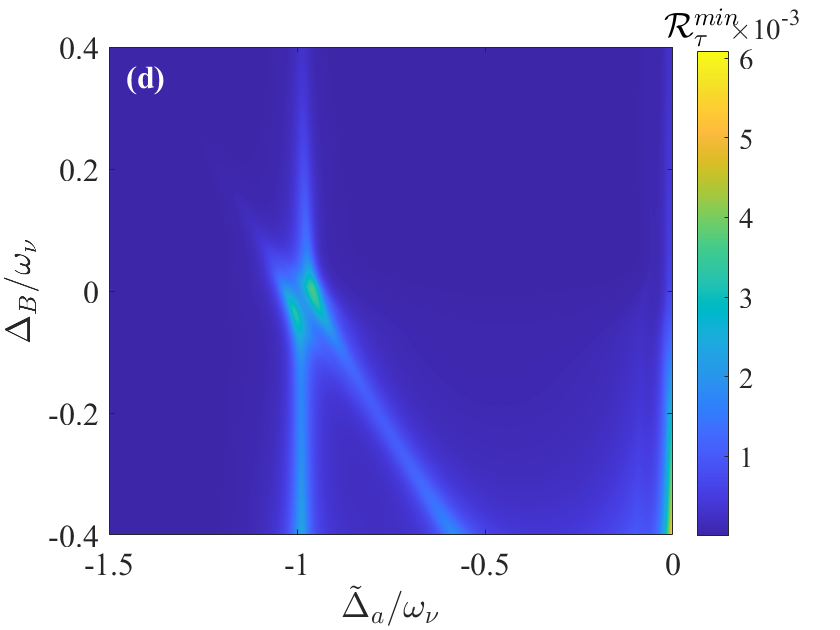}
	\caption{Contour plot of (a) bipartite entanglement for photon-magnon modes ($E_{am}$), (b)  bipartite entanglement for photon-vibration modes ($E_{aB}$), (c)  bipartite entanglement for magnon-vibration modes ($E_{mB}$), and (d) the minimum residual cotangle $\mathcal{R}_\tau^{min}$ as a function of the normalized cavity detuning $\tilde{\Delta}_a$ and the magnetic field direction. Common parameters for all subplots, $\omega_\nu/2\pi=30 $THz, $\mathcal{E}_l/\omega_\nu = 3.8$, $\gamma_\nu/\omega_\nu= \num{0.005}$, $g_\nu/\omega_\nu = \num{3.3e-6}$, $g_a=g_\nu$, $g_m=g_\nu$, $\Delta_m/\omega_\nu=1$, $\kappa_a/\omega_\nu=0.0166$, $\kappa_a= \kappa_m$, $J/\omega_\nu = \num{0.2}$, $N=\num{e7}$, $\theta=0$, $\chi/\omega_\nu=0$ and $T = \SI{210}{\kelvin}$.}
	\label{fig:fig1}
\end{figure}

\begin{figure}[tbh]
	\centering
	\includegraphics[width=4.2cm]{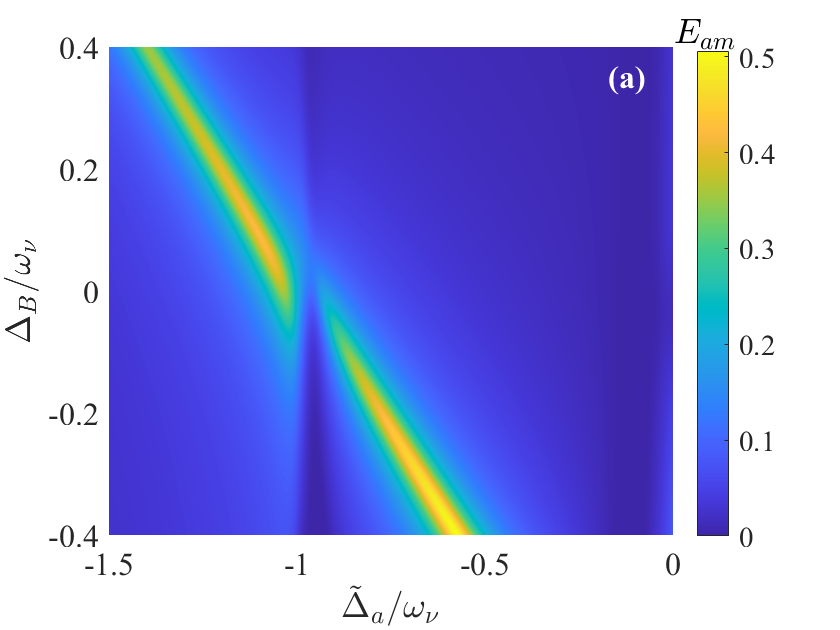}
	\includegraphics[width=4.2cm]{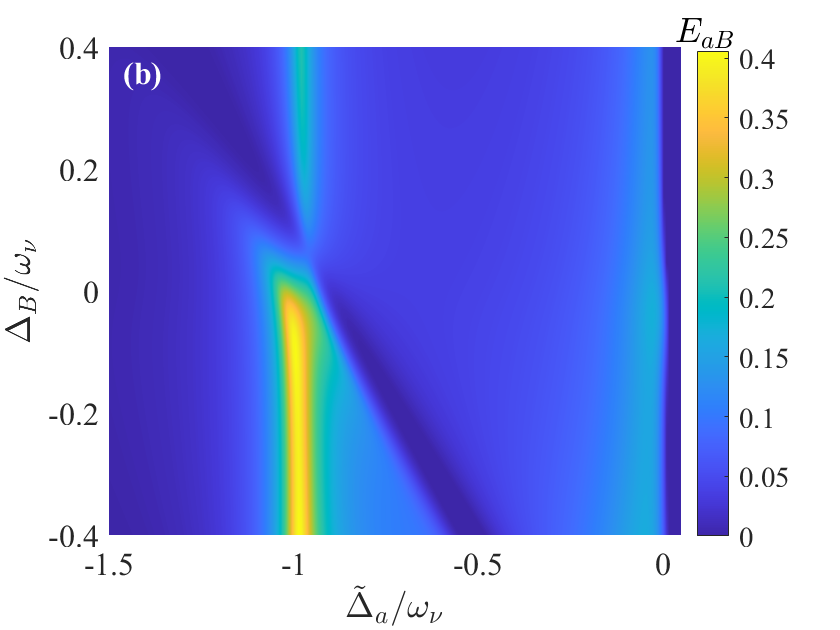}
	\includegraphics[width=4.2cm]{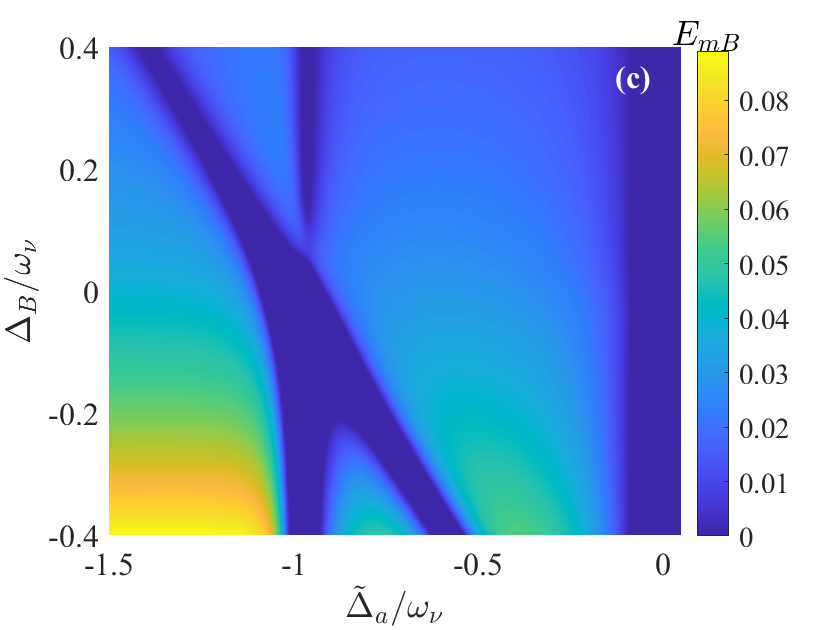}
	\includegraphics[width=4.2cm]{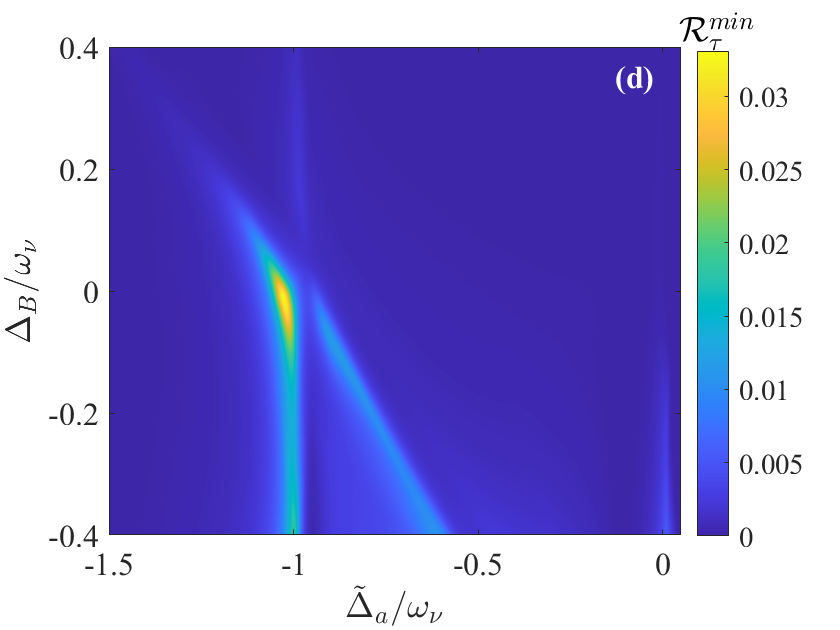}
	\caption{Contour plot of (a) bipartite entanglement for photon-magnon modes ($E_{am}$), (b)  bipartite entanglement for photon-vibration modes ($E_{aB}$), (c)  bipartite entanglement for magnon-vibration modes ($E_{mB}$), and (d) the minimum residual cotangle $\mathcal{R}_\tau^{min}$ as a function of the normalized cavity detuning $\tilde{\Delta}_a$ and the magnetic field direction. Common parameters for all subplots, $\omega_\nu/2\pi=30 $THz, $\mathcal{E}_l/\omega_\nu = 3.8$, $\gamma_\nu/\omega_\nu= \num{0.005}$, $g_\nu/\omega_\nu = \num{3.3e-6}$, $g_a=g_\nu$, $g_m=g_\nu$, $\Delta_m/\omega_\nu=1$, $\kappa_a/\omega_\nu=0.0166$, $\kappa_a= \kappa_m$, $J/\omega_\nu = \num{0.2}$, $N=\num{e7}$, $\theta=\pi/2$, $\chi/\omega_\nu=0.1$ and $T = \SI{210}{\kelvin}$.}
	\label{fig:fig2}
\end{figure}
In \Cref{fig:fig1}, we show the logarithmic negativity $E_{ij}$ and the minimum residual contangle $\mathcal{R}_\tau^{min}$ as a function of the normalized detuning $\tilde{\Delta}_{a}$ for different magnetic field direction, in the absence of magnon squeezing (i.e., when $\theta=0$, and $\chi/\omega_\nu=0$). From this panel of figures (\Cref{fig:fig1}), it is clear that the squeezing parameter $\chi$ has a significant effect on the  enhancement of bipartite and tripartite entanglements. In the absence of magnon squeezing, the generated bipartite and tripartite entanglements remain weak within the considered parameter regime. 

\Cref{fig:fig2} represents the logarithmic negativity $E_{ij}$ and the minimum residual contangle $\mathcal{R}_\tau^{min}$ as a function of the normalized detuning $\tilde{\Delta}_{a}$ for different magnetic field direction. As it can be seen, the photon-magnon bipartite entanglement, $E_{am}$, is generated when $-1<\tilde{\Delta}_a/\omega_\nu\leq-0.5$ for $-0.4\leq\Delta_m/\omega_\nu\leq 0.4$ (see \Cref{fig:fig2}(a)). One can also observe that the photon-vibration entanglement $E_{aB}$ peaks around anti-resonant condition i.e., $\tilde{\Delta}_a/\omega_\nu\sim-1$ as shown in \Cref{fig:fig2}(b). The underlying physical mechanism is that at resonance, the entanglement initially generated between magnons and vibrations, arising from magnetoelastic interactions is subsequently redistributed to the photon-magnon and photon-vibration subsystems. The bipartite entanglement for the magnon-vibration modes ($E_{mB}$) is diplayed in \Cref{fig:fig2}(c), where it can be seen that the entanglement vanishes around the anti-resonance, but it is generated away of that condition. Regarding the tripartite entanglement between the photon, magnon, and vibrational modes, it is depicted in \Cref{fig:fig2}(d), and it  emerges around anti-resonance. This indicates that, under this detuning condition, quantum correlations are not only present in the individual photon-magnon and photon-vibration pairs but also shared collectively among all three subsystems, forming genuine three-mode entanglement. 
\begin{figure}[tbh]
	\centering
	\includegraphics[width=4.2cm]{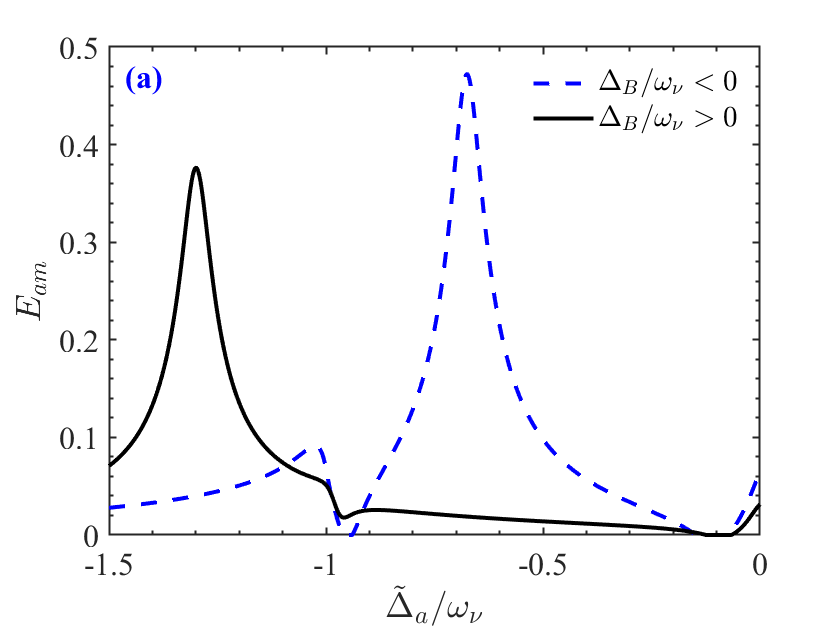}
	\includegraphics[width=4.2cm]{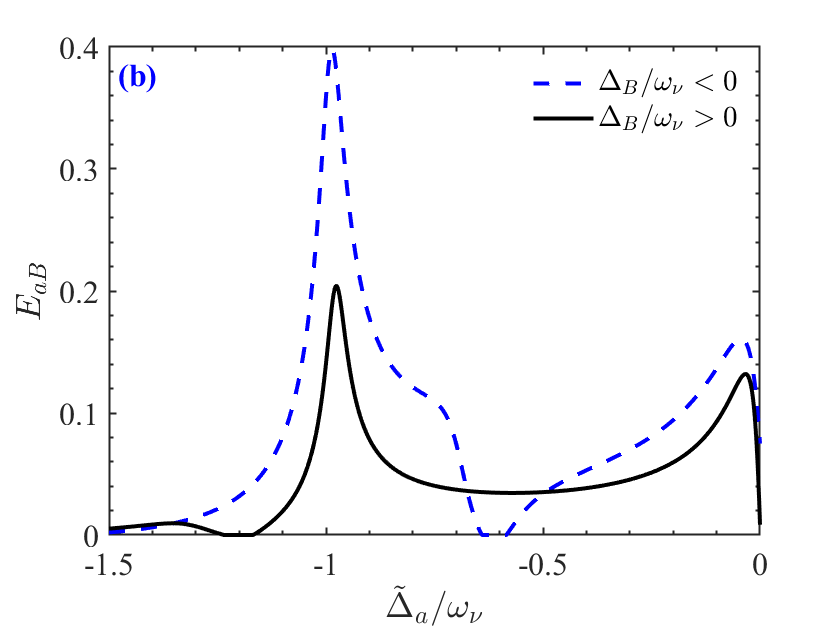}
	\includegraphics[width=4.2cm]{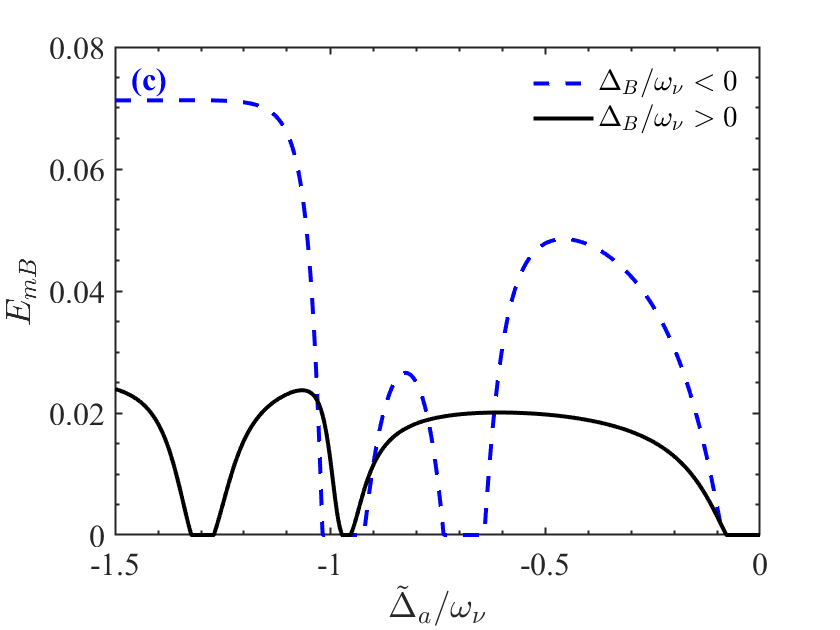}
	\includegraphics[width=4.2cm]{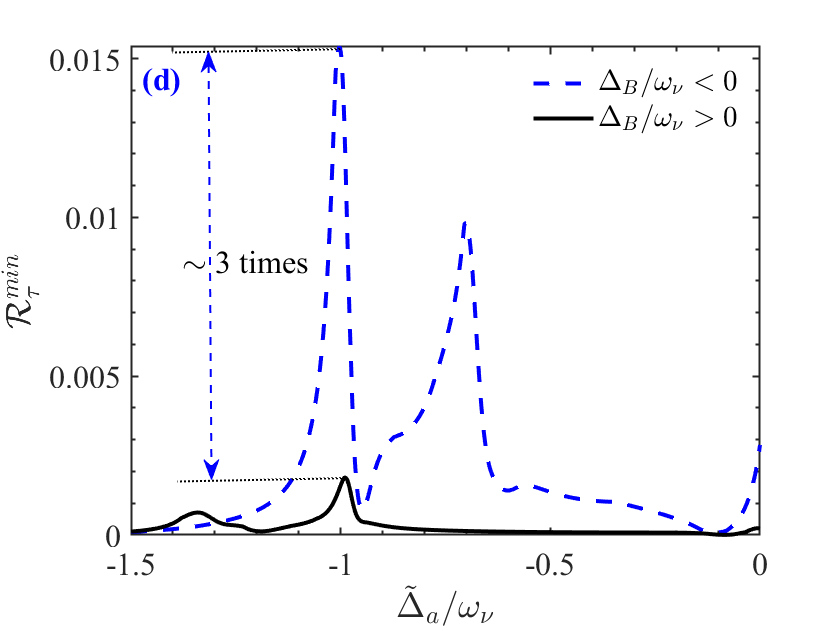}
	\caption{(a)  Plot of bipartite entanglement for photon-magnon modes ($E_{am}$), (b) entanglement for vibration-photon modes ($E_{aB}$), (c) entanglement for magnon-vibration modes ($E_{mB}$), and the (d) minimum residual cotangle $\mathcal{R}_\tau^{min}$, as a function of the normalized detuning $\tilde{\Delta}_{a}$. These plots are extracted from \Cref{fig:fig2}, for different direction of the magnetic field.  The common parameters for these figures are, $\omega_\nu/2\pi=30 $THz, $\mathcal{E}_l/\omega_\nu = 3.8$, $\gamma_\nu/\omega_\nu= \num{0.005}$, $g_\nu/\omega_\nu = \num{3.3e-6}$, $\tilde{\Delta}_m/\omega_\nu=1$, $|\Delta_B|/\omega_\nu=0.3$, $\kappa_a/\omega_\nu=0.0166$, $\kappa_a= \kappa_m$, $J/\omega_\nu = \num{0.2}$, $N=\num{7}$, $\theta=\pi/2$, $\chi/\omega_\nu=0.1$ and $T = \SI{210}{\kelvin}$.}
	\label{fig:fig3}
\end{figure}

In order to investigate nonreciprocal entanglement, we plot the three $E_{ij}$ logarithmic negativities and the minimum residual contangle $\mathcal{R}_\tau^{min}$ as functions of the normalized cavity frequency detuning $\tilde{\Delta}_a$ in \Cref{fig:fig3}. When the YIG sphere rotates in a fixed direction, a magnetic field applied along z(-z) produces a positive (negative) frequency shift $ \Delta_B$ via the Barnett effect, similar to the Sagnac effect. Therefore, the Barnett effect can be used to break the time-reversal symmetry and realize unidirectional bipartite and tripartite entanglement in our system (see \Cref{fig:fig3} for instance).
\begin{figure*}[tbh]
	\centering
	\includegraphics[width=5.5cm]{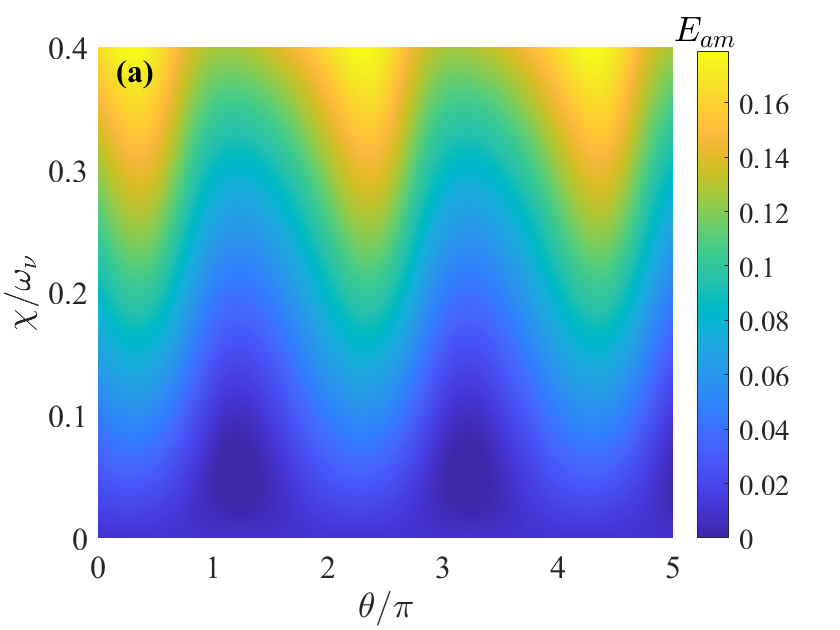}
	\includegraphics[width=5.5cm]{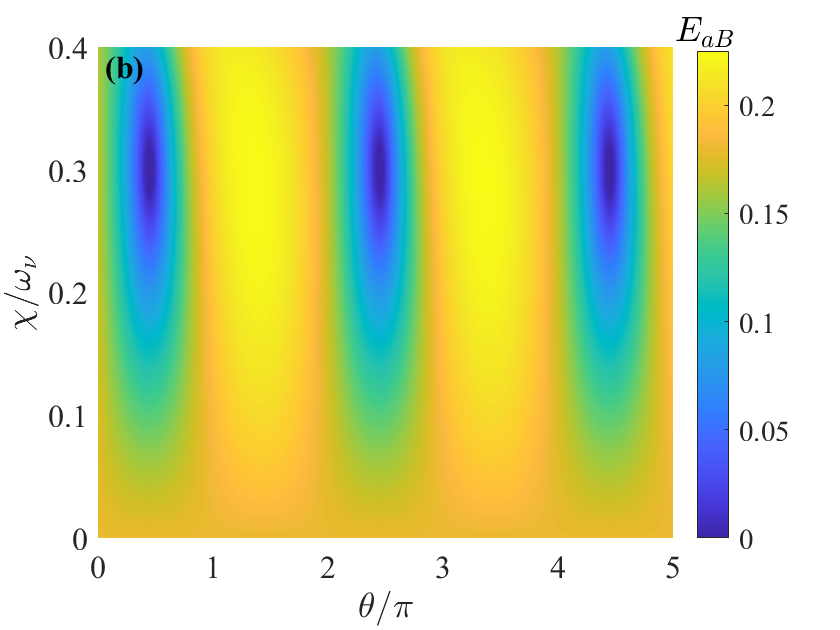}
	\includegraphics[width=5.5cm]{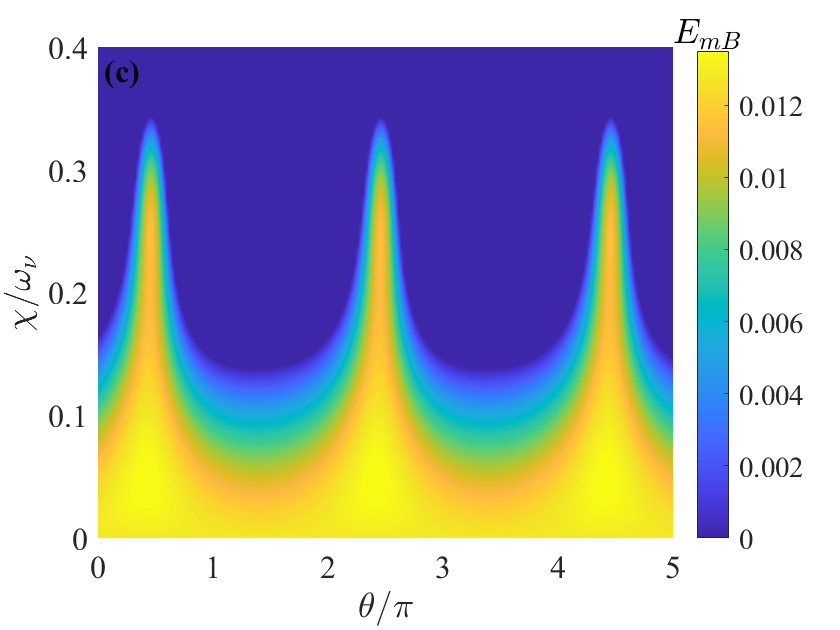}
	\caption{Contour plot of (a) bipartite entanglement for photon-magnon modes ($E_{am}$), (b)  bipartite entanglement for photon-vibration modes ($E_{aB}$), and (c)  bipartite entanglement for magnon-vibration modes ($E_{mB}$) as a function of phase $\theta$ for different values of squeezing parameter $\chi$. Common parameters for all subplots, $\omega_\nu/2\pi=30 $THz, $\mathcal{E}_l/\omega_\nu = 3.8$, $\gamma_\nu/\omega_\nu= \num{0.005}$, $g_\nu/\omega_\nu = \num{3.3e-6}$, $g_a=g_\nu$, $g_m=g_\nu$, $T = \SI{210}{\kelvin}$, $\Delta_m/\omega_\nu=1$, $|\Delta_B|/\omega_\nu=0.3$, $\kappa_a/\omega_\nu=0.0166$, $\kappa_a= \kappa_m$, $J/\omega_\nu = \num{0.2}$, $N=\num{e7}$, and $T = \SI{210}{\kelvin}$.}
	\label{fig:fig44}
\end{figure*}

 As shown in \Cref{fig:fig3}, when the Barnett shift is negative $\Delta_B<0$, both bipartite and tripartite entanglements are enhanced, whereas they are drastically suppressed when the Barnett shift is positive  $\Delta_B>0$, i.e., all the entanglements exhibit less responses when the direction of the magnetic field is reversed. This indicates that the sign of the Barnett-induced frequency shift plays a crucial role in controlling the strength of quantum correlations in the system, with negative shifts favouring stronger entanglement and positive shifts weakening it. This induces a threefold enhancement of the tripartite entanglement.
\begin{figure*}[tbh]
	\centering
	\includegraphics[width=5.5cm]{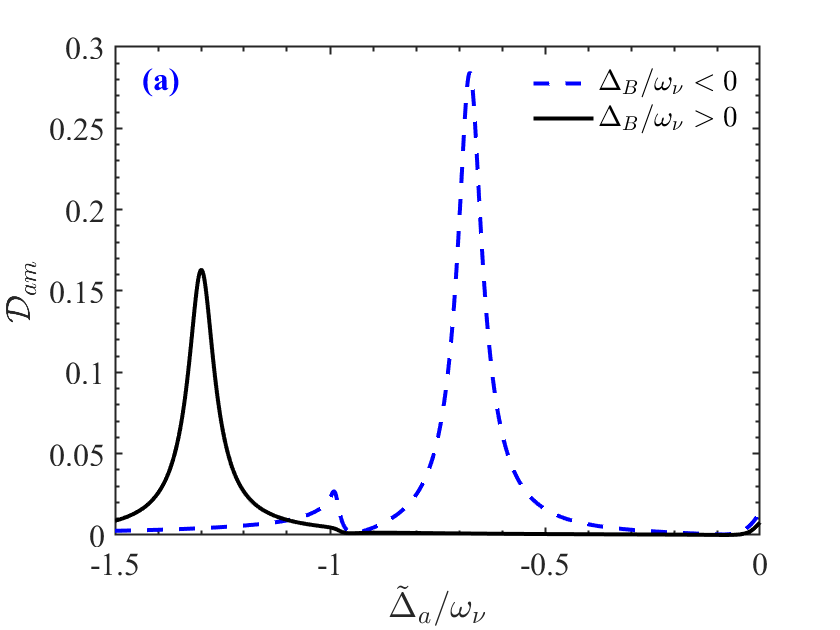}
	\includegraphics[width=5.5cm]{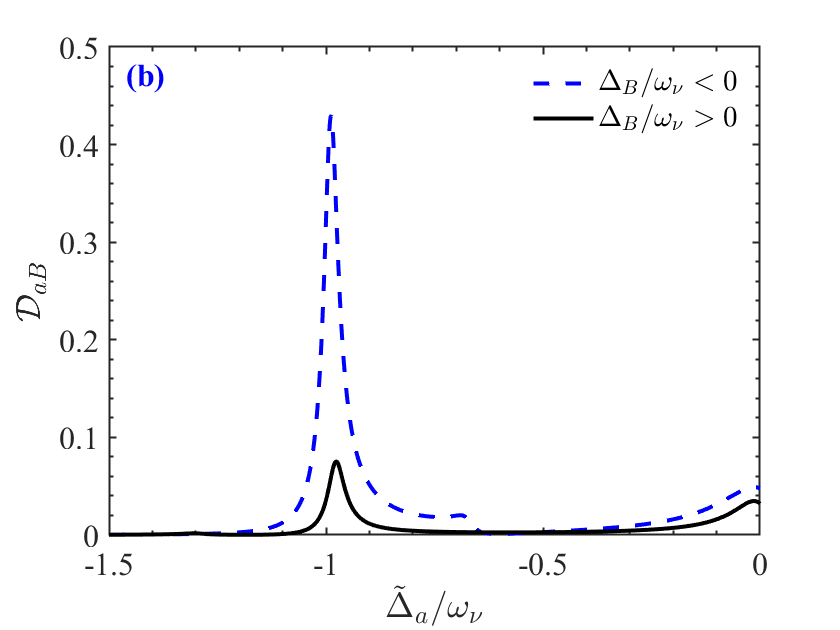}
	\includegraphics[width=5.5cm]{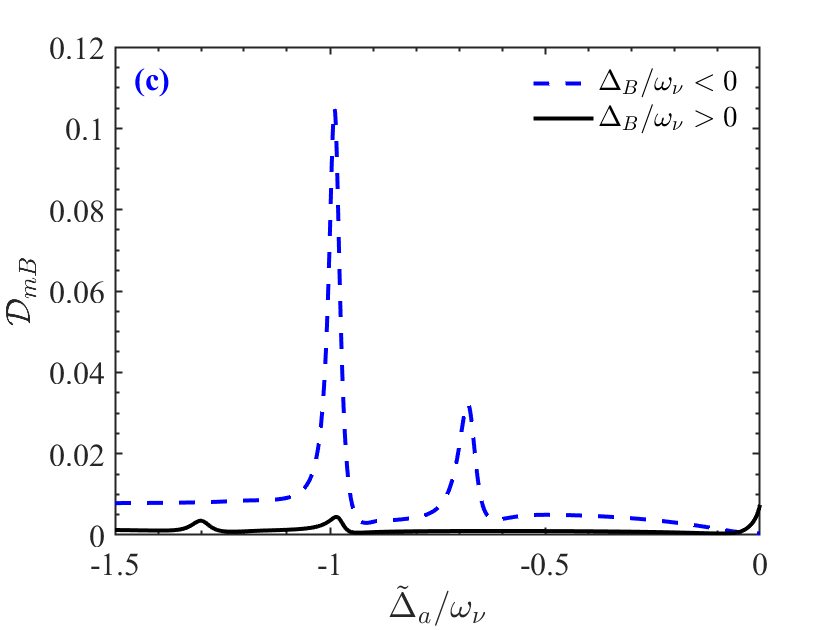}
	\caption{(a)  Plot of quantum discord for photon-magnon modes ($\mathcal{D}_{am}$). (b) Quantum discord for photon-vibration modes ($\mathcal{D}_{aB}$). (c) Quantum discord for magnon-vibration modes ($\mathcal{D}_{mB}$) as a function of detuning $\Delta_{a}$. The used parameters are, $\omega_\nu/2\pi=30 $THz, $\mathcal{E}_l/\omega_\nu = 3.8$, $\gamma_\nu/\omega_\nu= \num{0.005}$, $g_\nu/\omega_\nu = \num{3.3e-6}$, $\Delta_m/\omega_\nu=1$, $|\Delta_B|/\omega_\nu=0.3$, $\kappa_a/\omega_\nu=0.0166$, $N=\num{7}$, $\kappa_a= \kappa_m$, and $J/\omega_\nu = \num{0.2}$.}
	\label{fig:fig4}
\end{figure*}

In \Cref{fig:fig44}, we display the bipartite entanglement, $E_{am}$, $E_{aB}$ and $E_{mB}$ as a function of phase $\theta$ for different values of squeezing parameter $\chi$. It can be seen that all entanglements are significantly enhanced within a carefully chosen range of the phase $\theta$. Moreover, the degree of entanglement increases further with increasing squeezing parameter $\chi$, indicating the constructive role of squeezing of magnon in strengthening quantum correlations. In particular, the maximum photon-magnon, $E_{am}$, bipartite entanglement is achieved for phase values satisfying $\theta=\pi/2 + 2 n \pi$ where $ n\in \mathbb{Z}$. In contrary, the photon-vibration entanglement $E_{aB}$, reaches its maximum when the phase fulfils $\theta=3\pi/2 + 2n \pi$. This phase shift of  $\pi$ between the two optimal conditions highlights the complementary nature of the photon-magnon and photon-vibration entanglement generation mechanisms in the system. On the other hand, the magnon–vibration entanglement remains essentially independent of the phase in the weak magnon-squeezing regime $\chi/\omega_\nu<0.1$. As the squeezing strength increases beyond this value, the entanglement $E_{mB}$ becomes phase sensitive and reaches its maximum at $\theta=\pi/2 + 2 n \pi$ where $ n\in \mathbb{Z}$.

\begin{figure}[tbh]
	\centering
	\includegraphics[width=4.2cm]{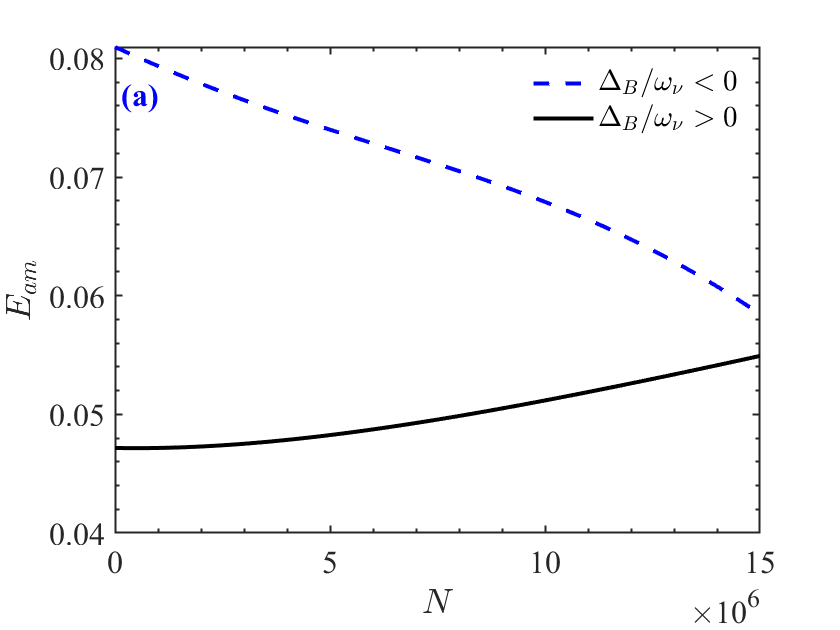}
	\includegraphics[width=4.2cm]{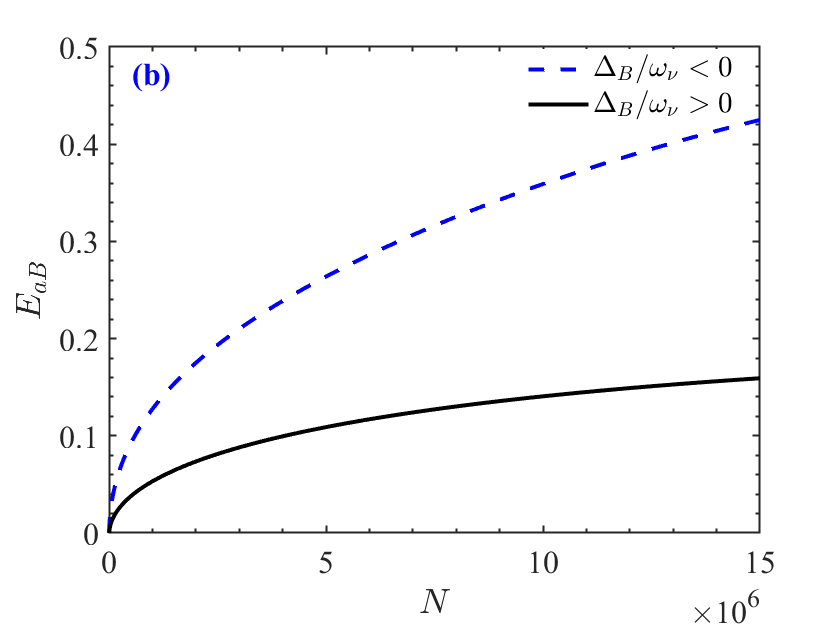}
	\includegraphics[width=4.2cm]{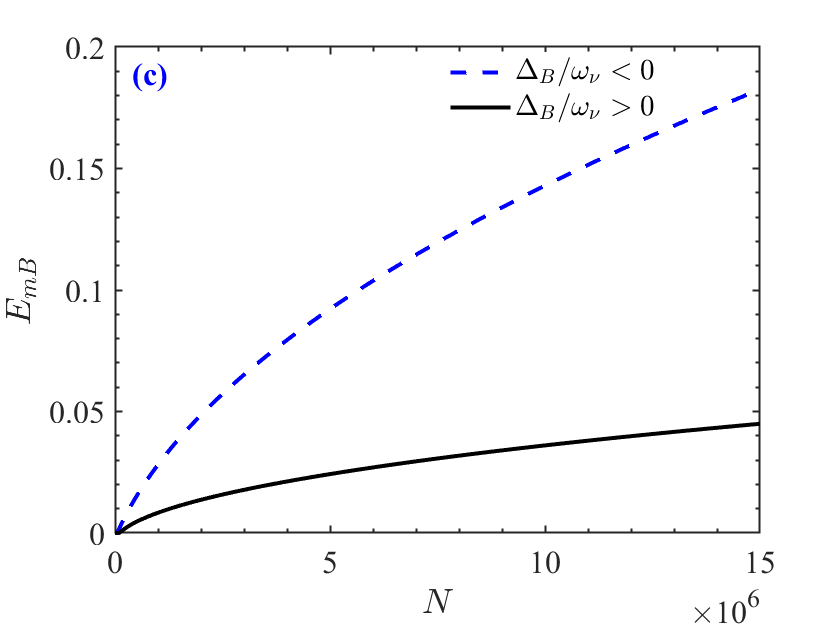}
	\includegraphics[width=4.2cm]{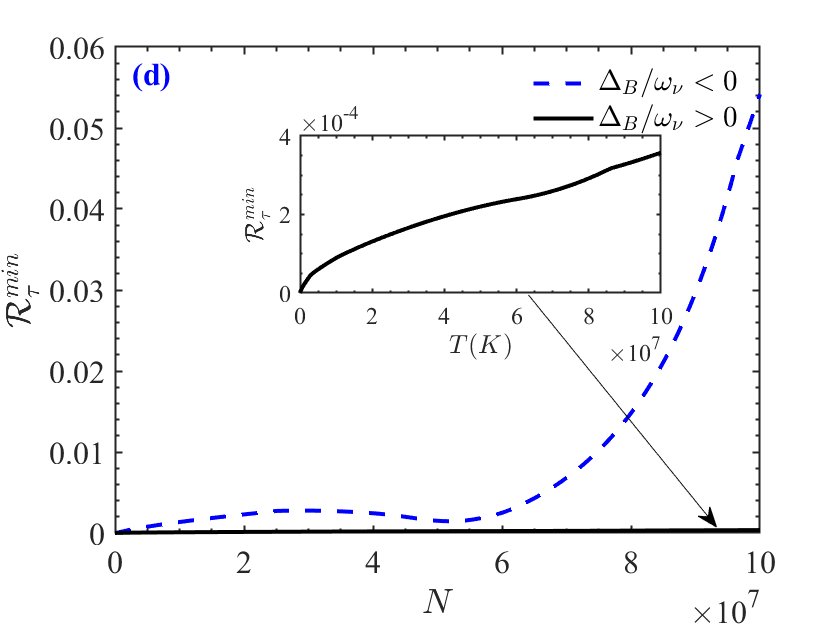}
	\caption{(a) Photon-magnon entanglement $E_{am}$ vs number of molecules N. (b) Photon-vibration entanglement $E_{aB}$ vs number of molecules N. (c) Entanglement for magnon-vibration modes ($E_{mB}$) with $\Delta_m/\omega_\nu=0.6$ and $\Delta_a/\omega_\nu=-0.5$ as a function of number of molecules N. (d) Minimum residual cotangle $\mathcal{R}^{min}$ with $\Delta_m/\omega_\nu=1$ and $\Delta_a/\omega_\nu=-0.5$ as a function of number of molecules N. The other used parameters are, $\omega_\nu/2\pi=30 $THz, $\mathcal{E}_l/\omega_\nu = 3.8$, $\gamma_\nu/\omega_\nu= \num{0.005}$, $g_\nu/\omega_\nu = \num{3.3e-6}$, $\tilde{\Delta}_a/\omega_\nu=-1$, $\Delta_m/\omega_\nu=1$, $|\Delta_B|/\omega_\nu=0.3$, $\kappa_a/\omega_\nu=0.0166$, $\kappa_a= \kappa_m$, $J/\omega_\nu = \num{0.2}$, $\theta=\pi/2$, $\chi/\omega_\nu=0.1$, and $T = \SI{210}{\kelvin}$.}
	\label{fig:fig6}
\end{figure}
In \Cref{fig:fig4} we plot the Gaussian Quantum Discord (GQD) as a function of the normalized cavity detuning $\tilde{\Delta}_a$. The GQD exhibits nonreciprocal behaviour similar to that of the entanglement, i.e., following the the direction of the applied magnetic field. Indeed, the plots in \Cref{fig:fig4} show how quantum correlation is strong when the Barnett frequency shift is negative ($\Delta_B<0$) and weak when it is positive ($\Delta_B>0$). Moreover, the photon-vibration and magnon-vibration quantum correlations are strong at the anti-resonance, i.e,. when the cavity detuning meets $\Delta_a/\omega_\nu\sim-1$. Among the different correlations, the magnon-vibration correlation is the weakest, whereas the photon-vibration correlation is the strongest. The underlying physics behind this observation is that, the magnetoelastic interaction does not benefit as strongly from the collective $\sqrt{N}$ enhancement as the optomechanical coupling. As a result, the magnon-vibration quantum discord remains relatively weak. In contrast, the photon-vibration and photon-magnon correlations are significantly enhanced due to cooperative couplings that scale with $\sqrt{N}$.

\Cref{fig:fig6} depicts the bipartite and tripartite entanglements as a function of molecular number $N$. The results reveal that all entanglements increase generally with number of molecules $N$ except for photon-magnon $E_{am}$. As it can be seen in \Cref{fig:fig6}(a), the photon-magnon $E_{am}$ entanglement decays with $N$ (for $\Delta_B<0$), but its value remains smaller than those of photon-vibration and magnon-vibration entanglements. The increase in photon-vibration and magnon-vibration entanglements arises because $E_{aB}$ and $E_{mB}$ are monotonically increasing function of the collective coupling strength $G_a=g_a\sqrt{N}$ and $G_m=g_m\sqrt{N}$. In contrast, the direct photon-magnon coupling $J$ does not benefit from such collective enhancement, limiting the magnitude of $E_{am}$  as $N$ increases.
\begin{figure}[tbh]
	\centering
	\includegraphics[width=4.2cm]{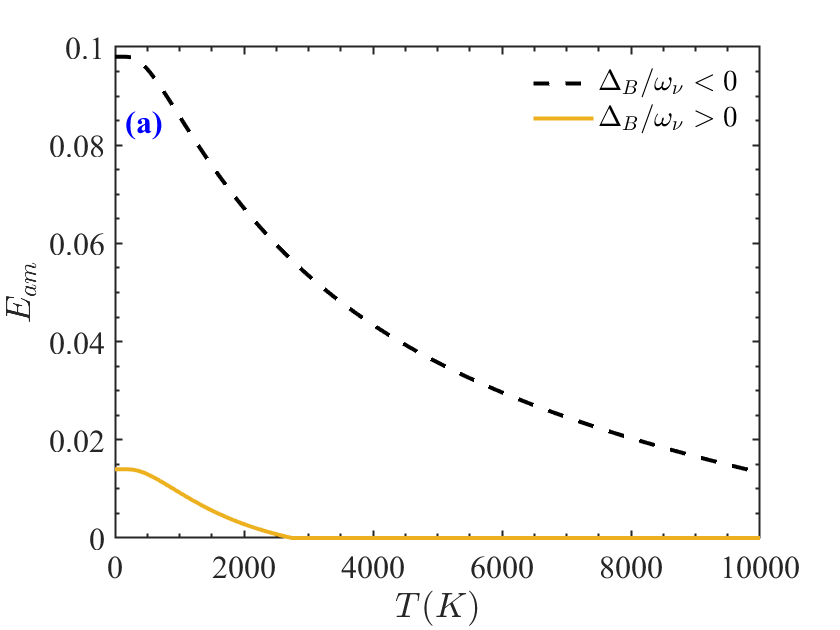}
	\includegraphics[width=4.2cm]{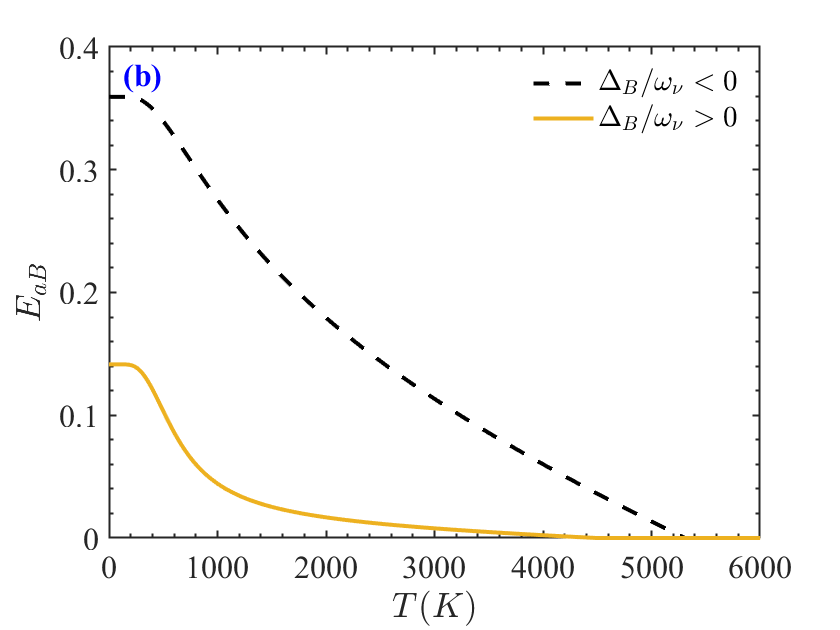}
	\includegraphics[width=4.2cm]{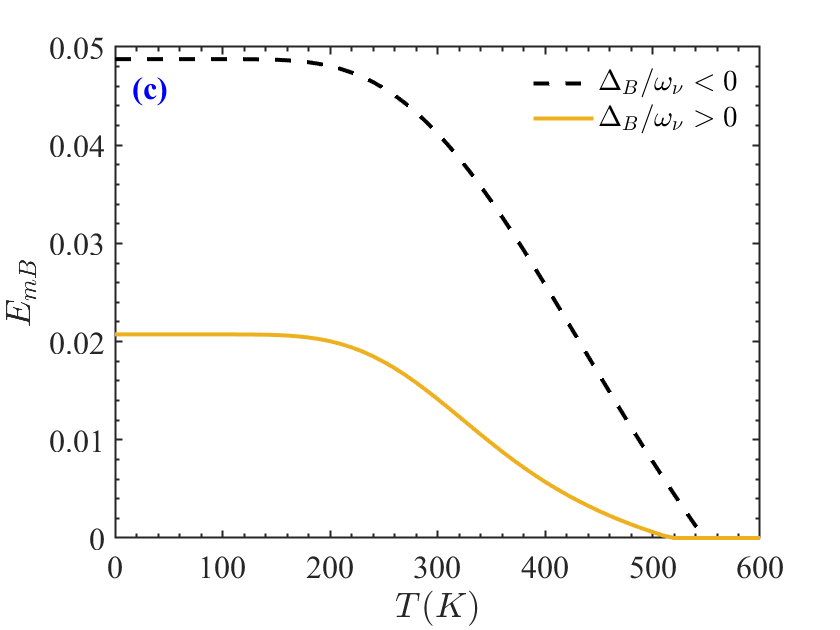}
	\includegraphics[width=4.2cm]{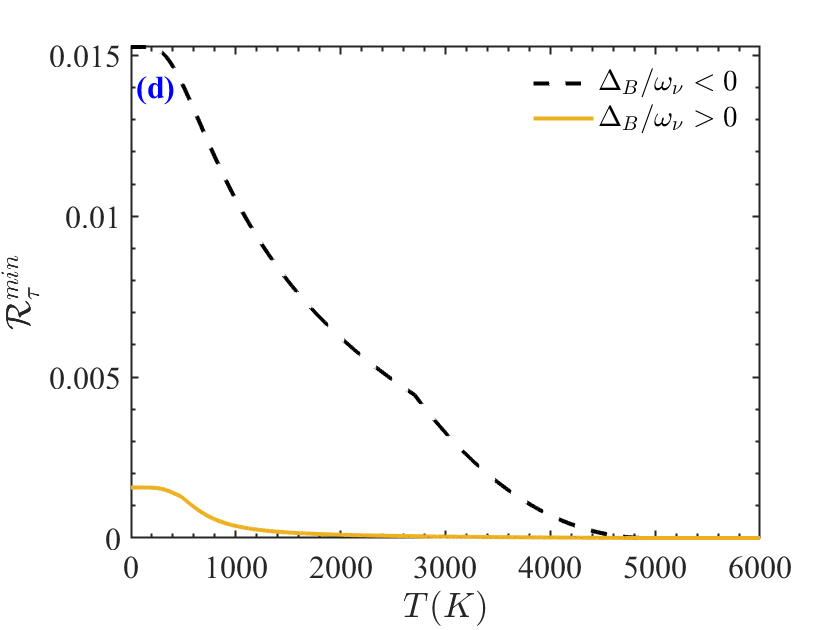}
	\caption{(a)  Plot of bipartite entanglement for photon-magnon modes ($E_{am}$) with $\Delta_a/\omega_\nu=-0.5$. (b) Entanglement for vibration-photon modes ($E_{aB}$). (c) Entanglement for magnon-vibration modes ($E_{mB}$) with $\Delta_m/\omega_\nu=1$ and $\Delta_a/\omega_\nu=-0.5$ as a function of temperature T. (d) Minimum residual cotangle $\mathcal{R}_\tau^{min}$ with $\Delta_m/\omega_\nu=1$ and $\Delta_a/\omega_\nu=-1$ as a function of temperature T. Common parameters for all subplots are, $\omega_\nu/2\pi=30 $THz, $\mathcal{E}_l/\omega_\nu = 3.8$, $\gamma_\nu/\omega_\nu= \num{0.005}$, $g_\nu/\omega_\nu = \num{3.3e-6}$, $\tilde{\Delta}_a/\omega_\nu=-1$, $\Delta_m/\omega_\nu=1$, $|\Delta_B|/\omega_\nu=0.3$, $\kappa_a/\omega_\nu=0.0166$, $\kappa_a= \kappa_m$, $J/\omega_\nu = \num{0.2}$, $N=\num{e7}$, $\theta=\pi/2$, $\chi/\omega_\nu=0.1$ and $T = \SI{210}{\kelvin}$.}
	\label{fig:fig7}
\end{figure}

In \Cref{fig:fig7}, we display the robustness of the bipartite and tripartite entanglements against temperature. Our findings strikingly unveil the highly thermal resilience of the proposed magnon-molecular hybrid system. Unlike conventional optomechanical and magnomechanical systems, which lose coherence rapidly at high temperatures, our generated quantum correlations persist at those temperatures. This remarkable resilience can be attributed to the high frequency vibrations of the molecules, which arise from their collective nature.  Consequently, entanglement remains largely unaffected by thermal noise, allowing the system to maintain coherent quantum behavior in regimes where conventional optomechanical and magnomechanical systems would be completely uncorrelated. This makes the magnon-molecular platform highly promising for  room-temperature quantum technologies.
\begin{figure*}[tbh]
	\centering
	\includegraphics[width=5.5cm]{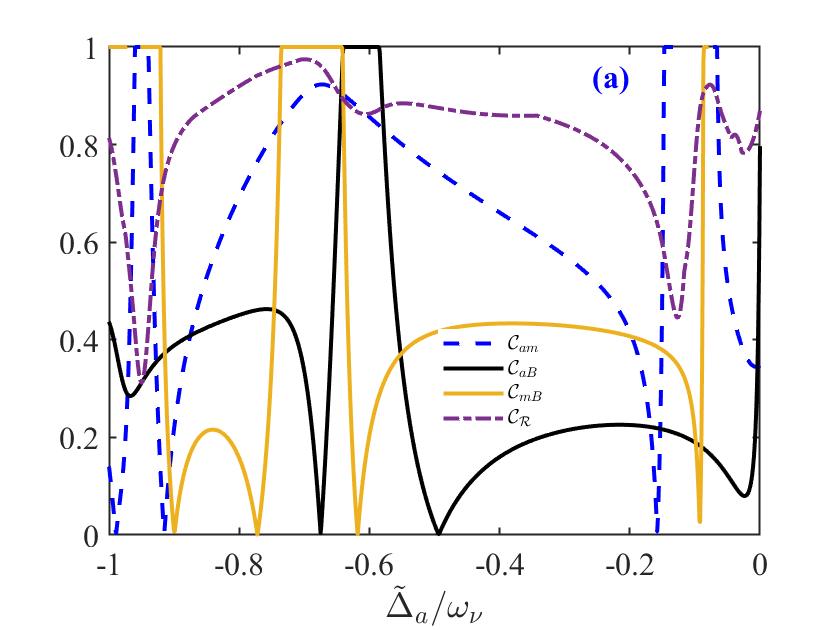}
	\includegraphics[width=5.5cm]{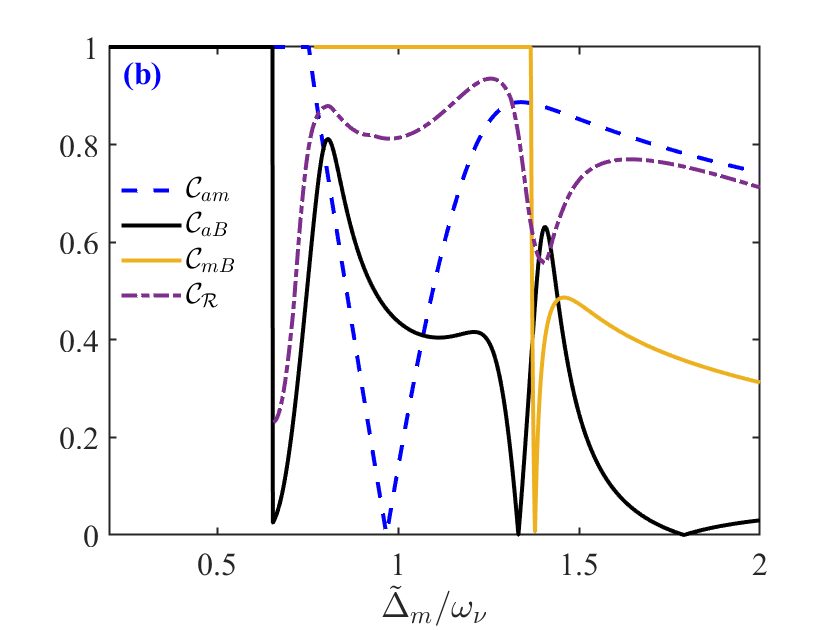}
	\includegraphics[width=5.5cm]{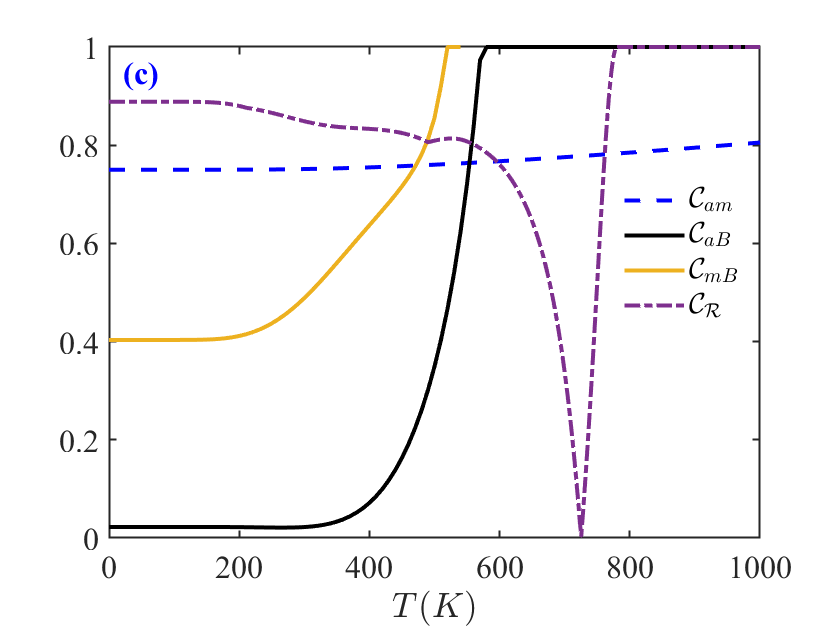}
	\caption{(a)  Plot of bidirectional contrast ratio $\mathcal{C}_{am}$, $\mathcal{C}_{aB}$, $\mathcal{C}_{mB}$ and $\mathcal{C}_\mathcal{R}$ as a function of cavity detuning $\tilde{\Delta}_{a}$. (b) bidirectional contrast ratio $\mathcal{C}_{am}$, $\mathcal{C}_{aB}$, $\mathcal{C}_{mB}$ and $\mathcal{C}_\mathcal{R}$ as a function of magnon detuning $\tilde{\Delta}_{m}$. (c) bidirectional contrast ratio $\mathcal{C}_{am}$, $\mathcal{C}_{aB}$, $\mathcal{C}_{mB}$ and $\mathcal{C}_\mathcal{R}$ for $\Delta_a/\omega_\nu=-0.5$, as a function of temperature T. The used parameters for these figures are, $\omega_\nu/2\pi=30 $THz, $\mathcal{E}_l/\omega_\nu = 3.8$, $\gamma_\nu/\omega_\nu= \num{0.005}$, $g_\nu/\omega_\nu = \num{3.3e-6}$, $\Delta_m/\omega_\nu=1$, $\tilde{\Delta}_a/\omega_\nu=1$, $\kappa_a/\omega_\nu=0.0166$, $\kappa_a= \kappa_m$, $J/\omega_\nu = \num{0.2}$, $N=\num{e7}$, $\theta=\pi/2$, $\chi/\omega_\nu=0.1$, and $T = \SI{210}{\kelvin}$.}
	\label{fig:fig9}
\end{figure*}

Furthermore, operating a molecular cavity optomechanical system at room-temperatures is highly challenging. The main difficulties arise from molecular instability, material degradation, and strong decoherence. Conventional optical cavity materials, such as gold and silver, are not stable at temperatures above approximately 500 K. At such temperatures, they can melt, recrystallize, or undergo morphological changes. Therefore, refractory materials are required. One promising candidate is titanium nitride, a plasmonic refractory material whose dielectric properties remain relatively stable at high temperatures and in harsh environments~\cite{Reddy2017}. However, even refractory metals can suffer from surface degradation due to oxidation and chemical reactions.
\subsection{Switchable nonreciprocity}{\label{sec:Bid}} 

Here, we use the following bidirectional contrast ratio $\mathcal{C}_{ij}$ ($\mathcal{C_R}$) for bipartite  (tripartite) entanglement to quantitatively describe nonreciprocal entanglement \cite{Chen2023}, 
\begin{equation}\label{eq:10}
\begin{aligned}
\mathcal{C}_{ij}&=\frac{|E_{ij}(\Delta_B>0)-E_{ij}(\Delta_B<0)|}{E_{ij}(\Delta_B>0)+E_{ij}(\Delta_B<0)},\\
\mathcal{C}_{\mathcal{R}}&=\frac{|\mathcal{R}^{min}_{\tau}(\Delta_B>0)-\mathcal{R}^{min}_{\tau}(\Delta_B<0)|}{\mathcal{R}^{min}_{\tau}(\Delta_B>0)+\mathcal{R}^{min}_{\tau}(\Delta_B<0)}.
\end{aligned}
\end{equation}

From these definitions, the bidirectional contrast ratio satisfies the condition $0\leq\mathcal{C}\leq1$. These bidirectional contrasts provide quantitative measure of nonreciprocal entanglement, i.e.,  the closer they appproach to unity, the more pronounced the entanglement is asymmetric between the interacting modes. We numerically plot the contrast ratio versus the normalized detuning $\tilde{\Delta}_a/\omega_\nu$ in \Cref{fig:fig9}(a) to clearly demonstrate this behaviour. It can be observed that the nonreciprocity of both bipartite and tripartite entanglements can be switched on or off by tuning the cavity detuning $\tilde{\Delta}_a/\omega_\nu$. Furthermore, the bidirectional contrast ratios for all entanglements can be tuned continuously from 0 to 1 by adjusting the cavity detuning $\tilde{\Delta}_a/\omega_\nu$. All entanglements in our system become perfectly nonreciprocal as the frequency detuning is adjusted. It can be also seen from \Cref{fig:fig9}(b) that perfect nonrecprocity can be achieved by varying the normalized magnon detuning $\tilde{\Delta}_m$. Our findings show that perfect nonreciprocity of photon-magnon entanglement can be achieved with our system for the temperature above $T>1000$K (see \Cref{fig:fig9}(c)). Similarly nonreciprocity of tripartite entanglement is achieved when $T>800 K$. These observations reveal how noise-tolerant quantum correlations can be engineered in our proposal. Such nonreciprocal robust quantum resources are useful for quantum processing, and plethora modern of quantum computational tasks.
 
\subsection{Experimental feasibility} 

The proposed magnon-molecular hybrid system can be realized experimentally, using currently available experimental techniques in cavity optomagnonics, cavity magnomechanics and molecular cavity quantum electrodynamics. To realize this setup, the magnon mode can be implemented using a YIG sphere placed inside a microwave cavity. Strong photon–magnon coupling with rates on the order of $\sim 90$ MHz has already been demonstrated experimentally~\cite{ bhoi2017, bittencourt2025}, enabling coherent magnon–photon interactions. The molecular ensemble can be introduced by embedding or coating a layer of molecules on or near the YIG sphere surface, or by placing them inside the same microwave cavity mode volume~\cite{Shalabney2015}. The molecular ensemble supports collective vibrational excitations that can couple to both the cavity field and the magnon mode via radiation pressure or magnetostrictive interactions. Since the effective coupling strengths scale as $G_{a,m}=g_{a,m}\sqrt{N}$, even moderate single-molecule coupling can lead to strong collective interactions for large (typically $\num{e6}-\num{e10}$ molecules), which is feasible using dense molecular films. The Barnett frequency shift $\Delta_B$ of  the magnon mode can be controlled by adjusting the rotation or magnetization of the YIG sphere. Previous experiments have demonstrated controlled mechanical rotation of YIG spheres by attaching them to rotors or air turbines, achieving angular frequencies of several kilohertz and even reaching gigahertz-scale rotation using levitated nanoparticles~\cite{Reimann2018,Schuck2018,Lu2025}. In our system, this tunable Barnett shift directly influences the direction dependent quantum correlations and can thus be used as a practical control knob. Potential experimental challenges here include maintaining low electromagnetic background noise in electrically driven systems and controlling temperature fluctuations in the rotating YIG sphere. However, experimental data show that frequency fluctuations on the order of $\pm30$ Hz at rotation frequencies of a few kilohertz are negligible, and the associated temperature variations ($\sim$0.1 K) are minimal~\cite{Barnett1915,Barnett2015,Barnett2009,Lu2025}.
Therefore, given the demonstrated capabilities in magnomechanical systems and the well-established control of molecular vibrational modes in cavity electrodynamics (QED), the realization of our proposed magnon-molecular system is experimentally feasible with current or near-future technology.

\section{Conclusion}\label{sec:concl}

In summary, we have investigated the generation of fundamental quantum correlations specifically entanglement, and quantum discord within a magnon-molecular system with magnon squeezing. Our findings demonstrate nonreciprocal entanglement, and quantum discord can be generated and control via the Barnett frequency shift $\Delta_B$. All the quantum corrections are enhanced when $\Delta_B<0$ and dramatically suppressed  when $\Delta_B>0$. Furthermore, by appropriately tuning the system parameters (e.g. phase $\theta$, squeezing parameter $\chi$, Barnett frequency shift $\Delta_B$, cavity detuning $\Delta_a$, magnon detuning $\Delta_m$ etc), perfect nonreciprocity of bipartite and tripartite entanglement can be achieved with our system. More interestingly, the generated entanglements demonstrate remarkable thermal robustness, remaining significant even at high temperatures. This exceptional resilience originates from the collective and high-frequency vibrational nature of the molecular ensemble, which leads to strong effective coupling strengths that dominate over thermal noise. Consequently, our results suggest that the proposed magnon-molecular system provides a promising platform for realizing high-temperature quantum information processing and quantum sensing, where robust and tunable nonreciprocal entanglement can be harnessed without the need for cryogenic cooling.

\section*{Acknowledgments}
P.D. acknowledges the Iso-Lomso Fellowship at Stellenbosch Institute for Advanced Study (STIAS), Wallenberg Research Centre at Stellenbosch University, Stellenbosch 7600, South Africa, and The Institute for Advanced Study, Wissenschaftskolleg zu Berlin, Wallotstrasse 19, 14193 Berlin, Germany. The research work was supported by Princess Nourah bint Abdulrahman University Researchers Supporting Project number (PNURSP2026R59), Princess Nourah bint Abdulrahman University, Riyadh, Saudi Arabia. The authors are thankful to the Deanship of Graduate Studies and Scientific Research at University of Bisha for supporting this work through the Fast-Track Research Support Program. 

\section*{Author Contributions}
 E.K.B. and P.D. conceptualized the work and carried out the simulations and analysis. A.-H. A.-A. and P.D.  participated in all the discussions and provided useful methodology and suggestions for the final version of the manuscript. N.A. and K.S.N. participated in the discussions and supervised the work. All authors participated equally in the writing, discussions, and the preparation of the final version of the manuscript.
\section*{Competing Interests} 
All authors declare no competing interests.
\section*{Data Availability}
Relevant data are included in the manuscript and supporting information. Supplementary data are available upon reasonable request.

\bibliography{Magnon}
\end{document}